\documentclass{article}
\usepackage{amssymb}
\usepackage{amsfonts}
\usepackage{amsmath}

\begin{document}

\title{Some remarks on relativistic zero-mass wave equations and
supersymmetry}
\author{Andrzej Okni\'{n}ski \\
Physics Division, Politechnika Swietokrzyska, \\
Al. 1000-lecia PP 7, 25-314 Kielce, Poland}
\maketitle

\begin{abstract}
We study several formulations of zero-mass relativistic equations, stressing
similarities between different frameworks. It is shown that all these
massless wave equations have fermionic as well as bosonic solutions.
\end{abstract}

\section{Introduction}

Observation that relativistic zero-mass equations may have fermionic as well
as bosonic solutions has a long history \cite%
{Foldy1956,Garbaczewski1986,Simulik1998,Bogush2009,Bialynicki2013}. More
exactly, it was shown that some zero-mas equations, for example the massless
Dirac equation, can describe fermionic as well as bosonic states. This
property, called the Fermi-Bose duality, was further studied for the massive
Dirac equation \cite{Simulik2011,Simulik2012}, see also \cite%
{Okninski2011,Okninski2012} where a notion of supersymmetry was invoked. The
most straightforward demonstration of the Fermi-Bose duality in the massless
case was given in \cite{Bialynicki2013}.

It is well known that zero-mass equations, e.g. Maxwell equations or the
Weyl neutrino equation, can be cast in several forms. In this note we study
relationships between three different formulations leading to these
equations. In the next Section we discuss spinor approach to zero-mass
equations \cite{Bialynicki2013}, then in Section 3 a relativistic equation
for a massless particle, involving $4\times 4$ matrix representation of the
Pauli algebra with a wavefunction with one zero component \cite%
{Borgardt1958,Moses1958,Lomont1958,Fushchych1983}, is described and in
Section 4 another approach to the Maxwell equations, based on the massless
Dirac equation \cite{Simulik1998}, is analysed. All these approaches are
more or less directly related to the Riemann--Silberstein vector \cite%
{Bogush2009,Bialynicki2013}. In Section 5 we study generalized Maxwell-like
solutions of these equations and in the last Section we discuss the obtained
results.

\newpage

\section{Spinor formulation of massless wave equations}

All relativistic massless equations can be uniformly written as \cite%
{Bialynicki2013}:%
\begin{equation}
\partial _{t}\varphi _{AB\ldots W}=-c\left( \mathbf{\sigma }\cdot \mathbf{%
\nabla }\right) _{A}^{\ Z}\varphi _{ZB\ldots W},  \label{zero-mass-gen}
\end{equation}%
where $\mathbf{\sigma }=$ $\left[ \sigma ^{1},\sigma ^{2},\sigma ^{3}\right] 
$ are the Pauli matrices.

Equation (\ref{zero-mass-gen}) for first -rank spinor $\varphi _{A}$ and
second-rank symmetric spinor $\varphi _{AB}$ leads to:%
\begin{eqnarray}
\partial _{t}\varphi _{A} &=&-c\left( \mathbf{\sigma }\cdot \mathbf{\nabla }%
\right) _{A}^{\ Z}\varphi _{Z},  \label{zero-mass-n} \\
\partial _{t}\varphi _{AB} &=&-c\left( \mathbf{\sigma }\cdot \mathbf{\nabla }%
\right) _{A}^{\ Z}\varphi _{ZB}\quad \left( \varphi _{AB}=\varphi
_{BA}\right) .  \label{zero-mass-M}
\end{eqnarray}

Equation (\ref{zero-mass-n}) is the Weyl neutrino equation while (\ref%
{zero-mass-M}) is equivalent to the Maxwell equations. To prove the
equivalence, the authors of Ref. (\cite{Bialynicki2013}) employed to their
Eqn. (12.3) well known relation between $\varphi _{AB}$ and the
Riemann-Silberstein vector $\mathbf{F}_{+}\mathbf{=E}+ic\mathbf{B}$:%
\begin{equation}
F_{+}^{1}=\phi _{11}-\phi _{00},\ F_{+}^{2}=-i\left( \phi _{11}+\phi
_{00}\right) ,\ F_{+}^{3}=2\phi _{01},  \label{RS}
\end{equation}%
where $\phi _{01}=\phi _{10}$.

Equation (\ref{zero-mass-M}) can be written in matrix form as (cf. Eqn
(12.3) of \cite{Bialynicki2013}):%
\begin{equation}
\partial _{0}\Psi =-c\mathbf{\Sigma }\cdot \mathbf{\nabla \,}\Psi ,
\label{zero-mass-nM1}
\end{equation}%
where%
\begin{equation}
\Sigma ^{1}=\left( 
\begin{array}{cccc}
0 & 0 & -1 & 0 \\ 
0 & 0 & 0 & -1 \\ 
-1 & 0 & 0 & 0 \\ 
0 & -1 & 0 & 0%
\end{array}%
\right) ,\ \Sigma ^{2}=\left( 
\begin{array}{cccc}
0 & 0 & i & 0 \\ 
0 & 0 & 0 & i \\ 
-i & 0 & 0 & 0 \\ 
0 & -i & 0 & 0%
\end{array}%
\right) ,\ \Sigma ^{3}=\left( 
\begin{array}{cccc}
-1 & 0 & 0 & 0 \\ 
0 & -1 & 0 & 0 \\ 
0 & 0 & 1 & 0 \\ 
0 & 0 & 0 & 1%
\end{array}%
\right) .  \label{Pauli1}
\end{equation}

Equation (\ref{zero-mass-nM1}) has Maxwell as well as neutrino solutions.
Indeed, if we choose in (\ref{zero-mass-nM1})$~\Psi _{M}\left( x\right)
=\left( \phi _{00},\ \phi _{01},\ \phi _{10},\ \phi _{11}\right) ^{T}$, $%
\phi _{01}=\phi _{10}$~then Eqn. (\ref{zero-mass-M}) is recovered while for $%
\Psi _{n}\left( x\right) =\varphi \otimes \xi \left( x\right) $ ($\otimes $
denotes the Kronecker product) with constant spinor $\varphi $ we get $%
\varphi \otimes \sigma ^{0}\partial _{0}\xi \left( x\right) =-c\varphi
\otimes \left( \sigma ^{1}\partial _{1}+\sigma ^{2}\partial _{2}+\sigma
^{3}\partial _{3}\right) \xi \left( x\right) $ where we used the obvious
representation of the matrices $\Sigma ^{i}$:%
\begin{equation}
\Sigma ^{1}=-\sigma ^{0}\otimes \sigma ^{1},\ \Sigma ^{2}=-\sigma
^{0}\otimes \sigma ^{2},\ \Sigma ^{3}=-\sigma ^{0}\otimes \sigma ^{3},
\label{Pauli2}
\end{equation}%
where $\sigma ^{0}=1_{2}$ is a $2\times 2$ unit matrix (matrices $\Sigma
^{i} $ are closely related to the Pauli matrices $\sigma _{i}$ what is
obvious from (\ref{Pauli2}) and, moreover, there are only three such
anticommuting matrices).

Moreover, there are also three matrices, $S^{1}=\sigma ^{1}\otimes \sigma
^{0}$, $S^{2}=\sigma ^{2}\otimes \sigma ^{0}$,$S^{3}=\sigma ^{3}\otimes
\sigma ^{0}$, commuting with $\Sigma ^{i}$'s. This result can be used to
project -- by application of operators $\frac{1}{2}\left( 1_{4}\pm
S^{3}\right) $ -- Eqn. (\ref{zero-mass-nM1}) and wavefunction $\Psi
_{n}\left( x\right) =\varphi \otimes \xi \left( x\right) $\ onto neutrino
subsolution $\xi \left( x\right) $ (note that $1_{4}\equiv \sigma
^{0}\otimes \sigma ^{0}$ is a $4\times 4$ unit matrix) to obtain the Weyl
neutrino equation%
\begin{equation}
\sigma ^{0}\partial _{0}\,\xi \left( x\right) =-c\left( \sigma ^{1}\partial
_{1}+\sigma ^{2}\partial _{2}+\sigma ^{3}\partial _{3}\right) \xi \left(
x\right) ,  \label{Weyl-n-1}
\end{equation}%
equivalent to (\ref{zero-mass-n}).

On the other hand, the wave function $\Psi _{M}=\left( \phi _{00},\ \phi
_{01},\ \phi _{10},\ \phi _{11}\right) ^{T}$, $\phi _{01}=\phi _{10}$, is an
eigenstate of the projection operator $R$:%
\begin{equation}
R=\left( 
\begin{array}{cccc}
1 & 0 & 0 & 0 \\ 
0 & \frac{1}{2} & \frac{1}{2} & 0 \\ 
0 & \frac{1}{2} & \frac{1}{2} & 0 \\ 
0 & 0 & 0 & 1%
\end{array}%
\right) ,  \label{R1}
\end{equation}%
i.e. $R\Psi _{M}=\Psi _{M}$ and thus we can write the Maxwell subsolution as:%
\begin{equation}
\partial _{0}R\Psi _{M}=-c\mathbf{\Sigma }\cdot \mathbf{\nabla \,}R\Psi _{M}.
\label{zero-mass-M1}
\end{equation}

From Eqn. (\ref{zero-mass-M1}) two equations follow:%
\begin{eqnarray}
\partial _{0}R\Psi _{M} &=&-c\left( R\mathbf{\Sigma }R\right) \cdot \mathbf{%
\nabla \,}R\Psi _{M},  \label{zero-mass-M2a} \\
0 &=&-c\left( 1_{4}-R\right) \mathbf{\Sigma }\cdot \mathbf{\nabla \,}R\Psi
_{M},  \label{zero-mass-M2b}
\end{eqnarray}%
which correspond to\ separation of the Maxwell equations (equation (\ref%
{zero-mass-M2b}) is equivalent to Eqn. (12.4) of Ref. \cite{Bialynicki2013}).

\section{Majorana-Oppenheimer formalism and massless wave equations}

According to Majorana and Oppenheimer the Maxwell equations can be written
in the following form (cf. \cite{Bogush2009} and references therein):

\begin{equation}
\partial _{0}\tilde{\Psi}=-c\mathbf{\tilde{\Sigma}}\cdot \mathbf{\nabla \,}%
\tilde{\Psi},  \label{zero-mass-nM2}
\end{equation}

where%
\begin{equation}
\tilde{\Sigma}^{1}=i\left( 
\begin{array}{cccc}
0 & -1 & 0 & 0 \\ 
1 & 0 & 0 & 0 \\ 
0 & 0 & 0 & 1 \\ 
0 & 0 & -1 & 0%
\end{array}%
\right) ,\ \tilde{\Sigma}^{2}=i\left( 
\begin{array}{cccc}
0 & 0 & -1 & 0 \\ 
0 & 0 & 0 & -1 \\ 
1 & 0 & 0 & 0 \\ 
0 & 1 & 0 & 0%
\end{array}%
\right) ,\ \tilde{\Sigma}^{3}=i\left( 
\begin{array}{cccc}
0 & 0 & 0 & -1 \\ 
0 & 0 & 1 & 0 \\ 
0 & -1 & 0 & 0 \\ 
1 & 0 & 0 & 0%
\end{array}%
\right) .  \label{Pauli3}
\end{equation}%
Matrices $\tilde{\Sigma}_{i}$ can be written as:%
\begin{equation}
\tilde{\Sigma}^{1}=\sigma ^{2}\otimes \sigma ^{3},\ \tilde{\Sigma}%
^{2}=\sigma ^{0}\otimes \sigma ^{2},\ \tilde{\Sigma}^{3}=\sigma ^{2}\otimes
\sigma ^{1}.  \label{Pauli4}
\end{equation}

It follows that matrices $\tilde{\Sigma}^{i}$'s are another $4\times 4$
representation of the Pauli matrices, analogous to representation (\ref%
{Pauli2}). More exactly, there are only three such anticommuting matrices
and there are also three matrices, $\tilde{S}^{1}=\sigma ^{2}\otimes \sigma
^{0}$, $\tilde{S}^{2}=\sigma ^{3}\otimes \sigma ^{2}$, $\tilde{S}^{3}=\sigma
^{1}\otimes \sigma ^{2}$, commuting with all $\tilde{\Sigma}^{i}$'s
(matrices $\tilde{\Sigma}^{i}$, $\tilde{S}^{j}$ are proportional to matrices 
$\alpha ^{i}$, $\beta ^{j}$ introduced in \cite{Bogush2009}).

Eqn. (\ref{zero-mass-nM2}) has fermionic as well as bosonic solution with
close analogy to (\ref{zero-mass-nM1}). Indeed, choosing $\tilde{\Psi}$ as $%
\tilde{\Psi}_{n}\left( x\right) =\varphi \otimes \eta \left( x\right) $
where $\varphi =\left( -i,\ 1\right) ^{T}$ is eigenvector of $\sigma ^{2}$, $%
\sigma ^{2}\varphi =+\varphi $, we arrive at equation:%
\begin{equation}
\left( \sigma ^{0}\partial _{0}+c\sigma ^{3}\partial _{1}+c\sigma
^{2}\partial _{2}+c\sigma ^{1}\partial _{3}\right) \eta \left( x\right) =0,
\label{zero-mass-n2}
\end{equation}%
which can be transformed by application of the unitary operator $U=\frac{1}{%
\sqrt{2}}\sigma ^{2}\left( \sigma ^{1}+\sigma ^{3}\right) $ to the neutino
equation (this property of Eqn. (\ref{zero-mass-nM2}) hasn't been noticed
before):%
\begin{equation}
\left( \sigma ^{0}\partial _{0}-c\sigma ^{1}\partial _{1}-c\sigma
^{2}\partial _{2}-c\sigma ^{3}\partial _{3}\right) U\eta \left( x\right) =0.
\label{Weyl-n-2}
\end{equation}

On the other hand, for $\tilde{\Psi}_{M}\left( x\right) =\left( 0,\
F_{+}^{1},\ F_{+}^{2},\ F_{+}^{3}\right) $, where $\mathbf{F}_{+}$ is again
the Riemann-Silberstein vector, Eqn. (\ref{zero-mass-nM2}) is equivalent to
the Maxwell equations \cite{Fushchych1983,Bogush2009}. We can thus write:%
\begin{equation}
\partial _{0}\tilde{R}\tilde{\Psi}_{M}=-c\mathbf{\tilde{\Sigma}}\cdot 
\mathbf{\nabla \,}\tilde{R}\tilde{\Psi}_{M},  \label{zero-mass-M3}
\end{equation}%
where%
\begin{equation}
\tilde{R}=\left( 
\begin{array}{cccc}
0 & 0 & 0 & 0 \\ 
0 & 1 & 0 & 0 \\ 
0 & 0 & 1 & 0 \\ 
0 & 0 & 0 & 1%
\end{array}%
\right) .  \label{R2}
\end{equation}

The following equations%
\begin{eqnarray}
\partial _{0}\tilde{R}\tilde{\Psi}_{M} &=&-c\left( P\mathbf{\tilde{\Sigma}}%
P\right) \cdot \mathbf{\nabla \,}\tilde{R}\tilde{\Psi}_{M},
\label{zero-mass-M4a} \\
0 &=&-c\left( 1_{4}-\tilde{R}\right) \mathbf{\tilde{\Sigma}}\cdot \mathbf{%
\nabla \,}\tilde{R}\tilde{\Psi}_{M},  \label{zero-mass-M4b}
\end{eqnarray}%
are equivalent to Eqns. (\ref{zero-mass-M2a}), (\ref{zero-mass-M2b}).

\section{Four-component massless Dirac equation: neutrino and Maxwell
solutions}

We shall discuss now fermionic and bosonic subsolutions of the massless
Dirac equation:

\begin{equation}
i\gamma ^{\mu }\partial _{\mu }\Psi =0,  \label{Dirac1}
\end{equation}%
where $\partial _{\mu }=\frac{\partial }{\partial x^{\mu }}$ and $x^{0}=ct$.
Projection operators $Q_{\pm }=\frac{1}{2}\left( 1_{4}\pm \gamma ^{5}\right) 
$, $\gamma _{5}=i\gamma _{0}\gamma _{1}\gamma _{2}\gamma _{3}$,\ in spinor
representation of the Dirac matrices%
\begin{equation}
\gamma _{0}=\left( 
\begin{array}{cc}
0_{2} & 1_{2} \\ 
1_{2} & 0_{2}%
\end{array}%
\right) ,\ \mathbf{\gamma }=\left( 
\begin{array}{cc}
0_{2} & -\mathbf{\sigma } \\ 
\mathbf{\sigma } & 0_{2}%
\end{array}%
\right) ,\ \gamma _{5}=\left( 
\begin{array}{cc}
\sigma ^{0} & 0_{2} \\ 
0_{2} & -\sigma ^{0}%
\end{array}%
\right) ,  \label{gammaSP}
\end{equation}%
where $0_{2}$ is a $2\times 2$ zero matrix, split (\ref{Dirac1}) into the
Weyl neutrino/antineutrino equations:%
\begin{eqnarray}
i\partial _{t}\xi &=&-c\mathbf{\sigma }\cdot \mathbf{\nabla }\xi ,
\label{Weyl1} \\
i\partial _{t}\eta &=&+c\mathbf{\sigma }\cdot \mathbf{\nabla }\eta .
\label{Weyl2}
\end{eqnarray}

The massless Dirac equation has also bosonic solutions. Simulik and Krivsky
demonstrated \cite{Simulik1998} that the following substitution:

\begin{equation}
\Psi _{M}=\left( iE_{3},\ iE_{1}-E_{2},\ -cB_{3},\ -icB_{2}-cB_{1}\right)
^{T},  \label{SK}
\end{equation}%
introduced into the Dirac equation (\ref{Dirac1}) converts it for standard
representation of the Dirac matrices (\ref{gammaST})%
\begin{equation}
\gamma _{0}=\left( 
\begin{array}{cc}
1_{2} & 0_{2} \\ 
0_{2} & -1_{2}%
\end{array}%
\right) ,\ \mathbf{\gamma }=\left( 
\begin{array}{cc}
0_{2} & \mathbf{\sigma } \\ 
-\mathbf{\sigma } & 0_{2}%
\end{array}%
\right) ,\ \gamma _{5}=\left( 
\begin{array}{cc}
0_{2} & 1_{2} \\ 
1_{2} & 0_{2}%
\end{array}%
\right) ,  \label{gammaST}
\end{equation}%
into the set of Maxwell equations (note that in \cite{Simulik1998}\
convention $\hslash =c=1$ was used). Let us also note that $\gamma ^{5}\Psi
_{M}$, where $\gamma _{5}=i\gamma _{0}\gamma _{1}\gamma _{2}\gamma _{3}$,
substituted for $\Psi $ in (\ref{Dirac1}) leads to the same Maxwell
equations.\textbf{\ }

\section{Generalized Maxwell-like equations}

It has been found that zero-mass Dirac equation leads to generalized Maxwell
equations. Indeed, substituting $\Psi =\Theta _{M}$ in (\ref{Dirac1}),%
\begin{equation}
\Theta _{M}=\left( iE^{3}-B^{0},\ iE^{1}-E^{2},\ iE^{0}-B^{3},\
-iB^{2}-B^{1}\right) ^{T},  \label{Gen1}
\end{equation}%
we get:%
\begin{equation}
\begin{array}{r}
c\mathbf{\nabla }\cdot \mathbf{E}=-\partial _{t}E_{0} \\ 
c\mathbf{\nabla }\cdot \mathbf{B}=-\partial _{t}B_{0} \\ 
\frac{1}{c}\partial _{t}\mathbf{E}-c\mathbf{\nabla }\times \mathbf{B}=-%
\mathbf{\nabla }E_{0} \\ 
\partial _{t}\mathbf{B+\nabla }\times \mathbf{E}=-c\mathbf{\nabla }B_{0}%
\end{array}
\label{GenMaxwell}
\end{equation}%
i.e. Maxwell equations with gradient-type sources \cite{Simulik1998}.

It is interesting that virtually the same equations arise from Eqns. (\ref%
{zero-mass-M1}), (\ref{zero-mass-M3}) if we remove some restrictions imposed
on the appropriate wavefunctions. Indeed, substituting $\Psi =\Phi _{M}$
into (\ref{zero-mass-M1}) with%
\begin{equation}
\begin{array}{l}
\Phi _{M}=\left( -\tfrac{1}{2}\zeta ^{1\dot{2}},\tfrac{1}{2}\zeta ^{1\dot{1}%
},-\tfrac{1}{2}\zeta ^{2\dot{2}},\tfrac{1}{2}\zeta ^{2\dot{1}}\right) ^{T}
\\ 
\zeta ^{1\dot{1}}=F_{+}^{3}+F_{+}^{0},\ \zeta ^{1\dot{2}%
}=F_{+}^{1}-iF_{+}^{2},\ \zeta ^{2\dot{1}}=F_{+}^{1}+iF_{+}^{2},\ \zeta ^{2%
\dot{2}}=-F_{+}^{3}+F_{+}^{0}%
\end{array}
\label{GenMaxwell-fi}
\end{equation}%
we get Eqns. (\ref{GenMaxwell}) while for $\Psi =\tilde{\Phi}_{M}$ inserted
into (\ref{zero-mass-M3}) where%
\begin{equation}
\tilde{\Phi}_{M}=\left( F_{+}^{0},\ F_{+}^{1},\ F_{+}^{2},\ F_{+}^{3}\right)
^{T},  \label{Gen3}
\end{equation}%
we obtain Eqns. (\ref{GenMaxwell}) with $E^{0}$ and $B^{0}$ interchanged. In
all three cases we recover the Maxwell equations demanding that $%
E^{0}=B^{0}=0$.

\newpage

\section{Discussion}

We have studied several formulations of zero-mass relativistic wave
equations. It turns out that all considered equations, (\ref{zero-mass-nM1}
), (\ref{zero-mass-nM2}), (\ref{Dirac1}), have fermionic (Weyl neutrino) as
well as bosonic (Maxwell) solutions. Equations (\ref{zero-mass-nM1}), (\ref%
{zero-mass-nM2}) have very similar structure, however existence of fermionic
solutions in the latter case wasn't noticed before. Note that the matrices $%
\Sigma _{i}$'s (as well as $\tilde{\Sigma}_{i}$'s) are a $4\times 4$
representation of the Pauli algebra and the corresponding equations are
analogues of the Weyl equation. It is interesting that equations (\ref%
{zero-mass-nM1}), (\ref{zero-mass-nM2}), (\ref{Dirac1}) lead to virtually
identical generalized Maxwell-like equations, discovered for Eqn. ( \ref%
{Dirac1}) in \cite{Simulik1998}. These Maxwell-like equations can be related
to the Dirac equation with non-zero mass (however for the case of a static
scalar potential only) thus extending the Fermi-Bose duality to the massive
case \cite{Simulik2002} (see also earlier ideas exposed in Refs. \cite%
{Sallhofer1978,Sallhofer1986}). The Fermi-Bose duality, referred to as
supersymmetry, was also studied in Refs. \cite{Okninski2011,Okninski2012},
placing earlier results \cite{Okninski1981,Okninski1982} in a new context.

Our work is related to other studies of massless equations. It was shown
that square of the Dirac operator is supersymmetric, containing fermion and
boson sectors, and this was used to study the zero-mass Dirac equation in the
interacting case in flat \cite{Horvathy1989} and curved space \cite%
{Comtet1995}. There is also a very interesting analogy between Fermi-Bose
duality discussed above and duality invariance characteristic for the
Maxwell field -- all free bosonic and fermionic gauge fields are invariant with respect 
to duality transformation \cite{Deser2005}.

\end{document}